\documentclass[twocolumn,prd,nofootinbib,aps,floats,floatfix,amsmath,amssymb,secnumarabic]{revtex4} %

\usepackage{graphicx}
\usepackage[usenames,dvipsnames]{color}
\usepackage{amsmath,amssymb}
\usepackage{slashed}
\usepackage[colorlinks,citecolor=blue]{hyperref}

\def\sfrac#1#2{{\textstyle{#1\over #2}}}
\newcommand{\be}{\begin{equation}}
\newcommand{\ee}{\end{equation}}
\newcommand{\ba}{\begin{array}}
\newcommand{\ea}{\end{array}}
\newcommand{\bea}{\begin{eqnarray}}
\newcommand{\eea}{\end{eqnarray}}

\newcommand{\DM}{{\rm\scriptscriptstyle DM}}
\newcommand{\SI}{{\rm\scriptscriptstyle SI}}

\begin{document}
\title{Improved Electroweak Phase Transition with\\ Subdominant Inert Doublet Dark Matter}
\author{James M.\ Cline}

\email{jcline@physics.mcgill.ca}
\affiliation{Department of Physics, McGill University, 
             3600 Rue University, Montr\'eal, Qu\'ebec, Canada H3A 2T8}
\author{Kimmo Kainulainen}
\email{kimmo.kainulainen@jyu.fi}
\affiliation{Department of Physics, P.O.Box 35 (YFL), 
             FIN-40014 University of Jyv\"askyl\"a, Finland}
\affiliation{Helsinki Institute of Physics, P.O. Box 64,
             FIN-00014 University of Helsinki, Finland}

\begin{abstract}
The inert doublet dark matter model has recently gained attention as a possible 
means of facilitating a strongly first order electroweak phase transition (EWPT), 
as needed for baryogenesis. We extend previous results by considering the regime 
where the DM is heavier than half the Higgs mass, and its relic density is
determined by annihilation into $W,Z$ and Higgs bosons.  We find a 
large natural region of parameter space where the EWPT is strongly first
order, while the lightest inert doublet state typically contributes only 
$0.1-3\%$ of the total dark matter.  Despite this small density,
its interactions with nucleons are strong enough to be directly
detectable given a factor of 5 improvement over the current
sensitivity of XENON100.  A 10\% decrease in the branching ratio for
Higgs decays to two photons is predicted.

\end{abstract}
\maketitle

{\bf Introduction.}  The idea that the matter-antimatter asymmetry was 
created during the electroweak phase transition (EWPT), by the process
known as electroweak baryogenesis (EWBG), is appealing because it is
perhaps the only proposed mechanism of baryogenesis that can be
directly tested at currently accessible energies in particle collider
experiments~\cite{Kuzmin:1985mm,reviews}. The subject has enjoyed a resurgence 
of interest since the advent of the Large Hadron Collider (LHC).  It is well-established
that new physics is needed to create a sufficiently strongly first-order phase
transition~\cite{Kajantie:1996mn,Rummukainen:1998as}, which is one of the requirements 
for baryogenesis in this framework. Supersymmetry was the most-studied 
option for EWBG during the LEP (Large Electron Positron collider)
era~\cite{Carena:1996wj,Espinosa:1996qw,Delepine:1996vn,Cline:1996cr,
Losada:1996ju,Laine:1996ms,Bodeker:1996pc,de Carlos:1997ru,
Cline:1998hy,Losada:1998at}, but since the LHC has so far not found 
any evidence for supersymmetry, nonsupersymmetric extensions have received more
attention recently.  Moreover, new physics is also required to explain
the dark matter (DM) of the universe, so it is natural to inquire
whether the same ingredients might explain both DM and EWBG 
\cite{Dimopoulos:1990ai,Barger:2008jx,Kang:2009rd,Kumar:2011np,
Chung:2011it,Kozaczuk:2011vr,Carena:2011jy,Ahriche:2012ei,Espinosa:2011eu,
Espinosa:2011ax,Chowdhury:2011ga}.

Recently there has been interest \cite{Espinosa:2011eu,
Espinosa:2011ax,Chowdhury:2011ga,Borah:2012pu,
Gil:2012ya,Ahriche:2012ei,Cline:2012hg} in scalar 
dark matter models with respect to their potential for enhancing the
electroweak phase transition strength.  In the case where the scalar
$S$ is an electroweak doublet, 
it is ``inert'' due to the $Z_2$ symmetry $S\to -S$ that
prevents it from coupling to fermions, and which stabilizes the lowest mass
component against decay. 
Its couplings to the usual Higgs, of the schematic form $|H|^2|S|^2$,
can give rise to a strengthened EWPT as desired for EWBG. However in
the examples of the inert doublet model studied so far, fine-tuning of 
the dark matter mass with respect to the Higgs boson mass, $m_\DM \cong m_h/2$, was needed
in order to have a strong enough annihilation cross section for
achieving the right relic density
(via resonantly enhanced by Higgs exchange in the $s$-channel), while simultaneously getting a
strong enough phase transition~\cite{Borah:2012pu}. Moreover in these cases 
a second fine-tuning between different $|H|^2|S|^2$-type quartic couplings 
was needed, to keep the effective DM-Higgs interaction small enough to
satisfy constraints from direct detection experiments, notably
XENON100.

In the present work, we show that the parameter space for a strong
EWPT plus potentially detectable inert doublet DM is significantly
enlarged when we take into account heavier DM annihilations into $W$,
$Z$ and Higgs bosons, and if we allow for it to be a subdominant
component of the full DM population.  For these new examples, the
fine-tuning problems mentioned above are eliminated, making this a
more natural scenario.  Even though inert doublet DM may constitute only
0.1\% of the total dark matter in these examples, its interactions
with nuclei, mediated by the Higgs, can be sufficiently strong to make it
detectable with only modest improvements relative to the current
sensitivity of XENON-like experiments.   Such improvements are
expected in the near future, making this proposal highly testable.\\

{\bf Model and Methodology.}  The potential of the inert doublet model is taken to be
\begin{eqnarray}
  V &=& \frac{\lambda}{4}\left(H^{\dagger i}H_i 
         -\frac{v^2}{2}\right)^2 + m^2_1 (S^{\dagger i}S_i)\nonumber\\
    &+& \lambda_1 (H^{\dagger i}H_i)(S^{\dagger j}S_j)+\lambda_2 
        (H^{\dagger i}H_j)(S^{\dagger j}S_i) \nonumber \\
    &+&[\lambda_3 H^{\dagger i}H^{\dagger j} S_iS_j +\text{h.c.}] + 
\lambda_S (S^{\dagger i}S_i)^2
\end{eqnarray}
The combination of couplings relevant to the dark matter candidate (by convention taken 
to be the CP-even neutral member $H_0$ of the $S$ doublet) is
\be
	\lambda_\DM = \lambda_1+\lambda_2 + 2\lambda_3
\label{ldm}
\ee
It controls both the DM mass, $m^2_\DM = m_1^2 + \sfrac12\lambda_\DM v^2$, where 
$v=246$ GeV is the Higgs VEV, and the coupling to the real Higgs field $h$, 
$\sfrac14\lambda_\DM h^2s^2$.  In the following we will take the one-loop correction 
to the zero-temperature effective potential to be augmented by counterterms that preserve 
these tree-level relations.

\begin{figure}[tb]
\centerline{$\!\!\!$
\includegraphics[width=0.50\textwidth]{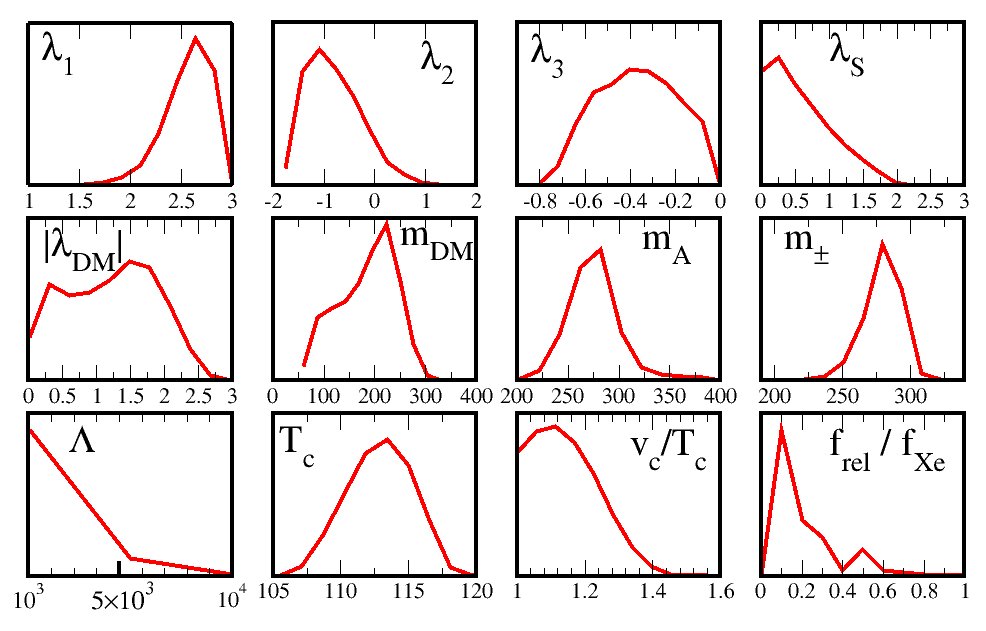}}
\vskip-0.3cm
\caption{Distributions of model parameters (both input and derived)
from Monte Carlo.
Dimensionful parameters are in units of GeV.}
\label{fig-hist}
\end{figure}

We use the same one-loop effective potential at finite temperature as
described in ref.\ \cite{Borah:2012pu} for the analysis of the phase
transition.  We also apply the same accelerator mass, vacuum stability and
electroweak precision data constraints as decribed there, as well as
fixing the Higgs mass to be 126 GeV~\cite{ATLAS:2012ae,Chatrchyan:2012tx,Chatrchyan:2012twa,ATLAS:2012ad}.  
Where we depart from ref.\
\cite{Borah:2012pu} is in our treatment of the annihilation cross
section that determines the $H_0$ relic density.  In ref.\ \cite{Borah:2012pu}, 
only annihilations to standard model fermions up to the $b$ quark were included.  
Here we use a similar expression as in ref.\ \cite{Cline:2012hg}, that includes 
more possible channels, notably into $W$ and $Z$ gauge bosons and higgs bosons, 
whose importance for DM-analysis was first observed in ref.~\cite{Enqvist:1988we}. 
We further extend the range of possible DM masses here to include annihilation 
into top quarks.

The annihilation cross section $\langle\sigma v\rangle$ is used to determine the relic density
by comparing to results of ref.\ \cite{Steigman:2012nb}, where the
value $\langle\sigma v\rangle_0$ that produces the full relic density,
as a function of dark matter mass, was accurately computed.  We define
the fraction of the full relic density attributable to the inert
doublet component as $f_{\rm rel} = {\langle\sigma v\rangle_0/ \langle\sigma v\rangle}$.
We define another ratio $f_{\rm Xe} = \sigma_{\rm Xe}/\sigma_\SI$ which 
involves the cross section for inert doublet scattering on nucleons,
$\sigma_\SI$, as given in ref.~\cite{Barbieri:2006dq} (see ref.~\cite{Cline:2012hg} 
for details of the Higgs coupling to nucleons), and the 
2012 limit of XENON100~\cite{Aprile:2012nq} $\sigma_{\rm Xe}$.  If the IDM scalar constituted
100\% of the dark matter, the XENON100 limit would be that $f_{\rm Xe}>1$.
But due to the subdominant density of this component, the constraint
is relaxed to $f_{\rm rel} < f_{\rm Xe}$. We demand that this be satisfied, in 
addition to the relic density constraint $f_{\rm rel}<1$. \\

\begin{figure}[tb]
\centerline{
\includegraphics[width=0.35\textwidth]{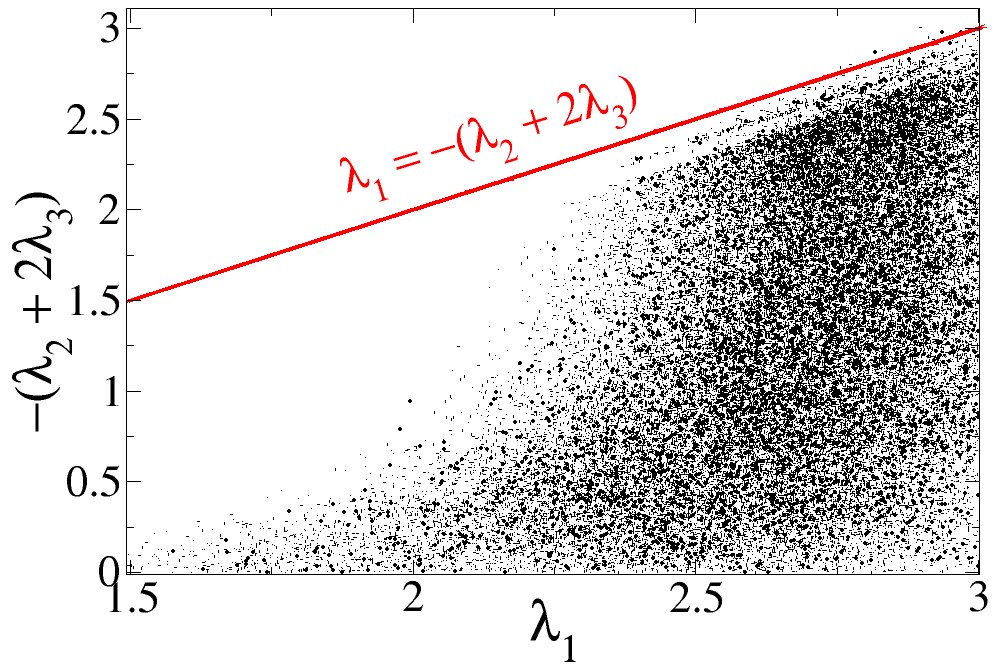}}
\vskip-0.3cm
\caption{Correlation between $\lambda_2+2\lambda_3$ and $\lambda_1$,
showing the absence of fine tuning $\lambda_1=-(\lambda_2+2\lambda_3)$
which would be present if $\lambda_{\DM}$ was required to be small.}
\label{fig-tuning}
\end{figure}

{\bf Monte Carlo results.}  We generated tens of thousands of models
using a Markov Chain Monte 
Carlo (MCMC) procedure in which 
 $(v_c/T_c)/\lambda_\DM$ was the quantity chosen to be biased toward
larger values, thus favoring both stronger phase transitions and
smaller $\lambda_\DM$ at the same time.  The latter criterion was not
necessary for our purposes, since ultimately small values of $\lambda_\DM$ 
were still disfavored, but it was instructive to try to enhance the
probability of such models in order to better illustrate the contrast
between the large-$\lambda_\DM$ and small-$\lambda_\DM$ regimes.
In previous studies, small values of $\lambda_\DM$ were required in
order to sufficiently suppress the direct detection cross section.
In the present work we find that larger values of $\lambda_\DM$ can
be consistent with XENON constraints due to the suppressed relic 
density.

Distributions of parameters resulting from the Monte Carlo run are  
plotted in Fig.~\ref{fig-hist}.  These include the quartic couplings  
of the potential, $\lambda_\DM$, the
three scalar masses $m_\DM$, $m_A$ and $m_\pm$ of the inert  doublet,
the scale $\Lambda$ of the nearest Landau pole from runing of the
renormalization group equations, the critical temperature $T_c$, the
measure $v_c/T_c$ of the phase transition strength, and the  ratio
$f_{\rm rel}/f_{\rm Xe}$ that shows how close the model comes to saturating the
XENON100 direct detection limit.  
It is seen that the magnitudes of the quartic
couplings are typically $O(1)$ to get a strong phase transition, 
as expected.  But in contrast to previous studies, there need no
longer be a conspiracy amongst large couplings to make the combination
$\lambda_\DM$ in eq.\
(\ref{ldm}) small.  This absence of tuning is further demonstrated by
Fig.~\ref{fig-tuning}, that shows most of the accepted models are not
clustered around the relation $\lambda_1 = -(\lambda_2 + 2\lambda_3)$
as was the case in ref.\ \cite{Borah:2012pu}.

\begin{figure*}[tb]
\centerline{$\!\!\!\!\!\!\!\!$
\includegraphics[width=0.32\textwidth]{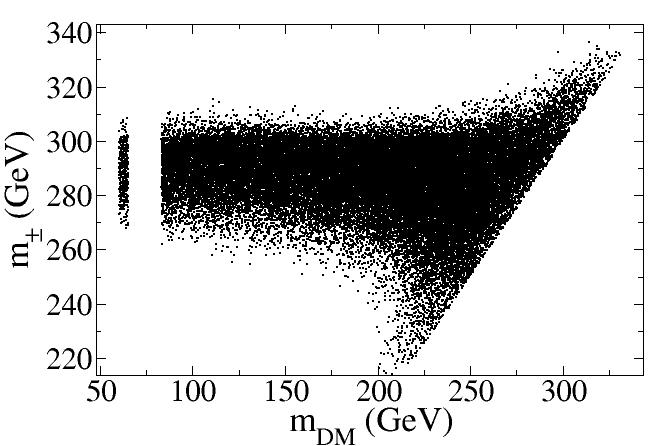}
\includegraphics[width=0.32\textwidth]{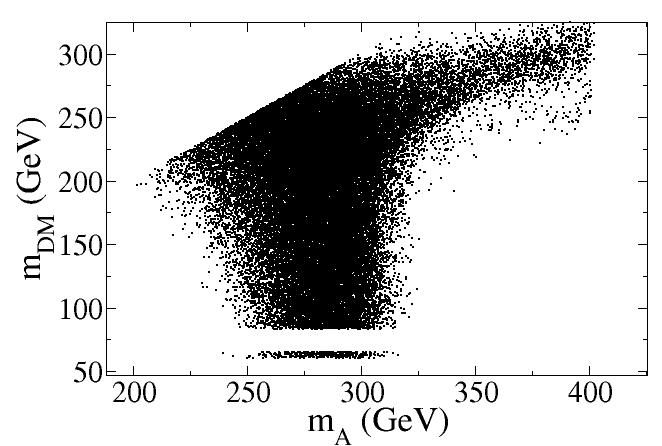}
\includegraphics[width=0.33\textwidth]{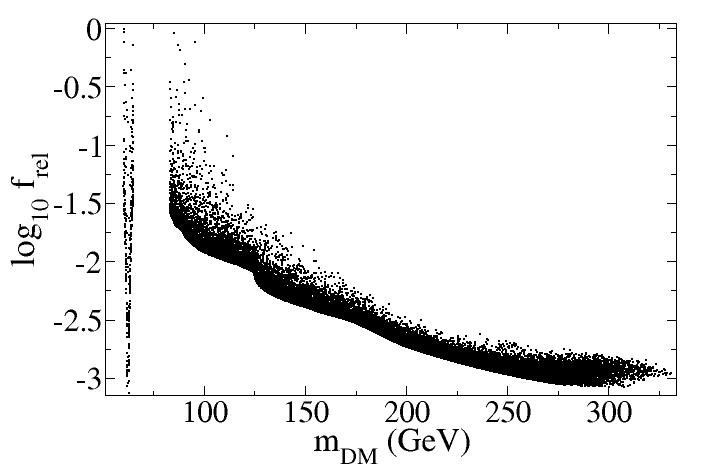}}
\vskip-0.4cm
\caption{Panels from the left to right: the correlation between the dark matter mass $m_{\rm DM}$ and the charged scalar mass $m_\pm$ (1), between $m_{\rm DM}$ and the CP-odd neutral scalar mass $m_A$ (2) and between $m_{\rm DM}$ and the logarithm of the fraction of the full relic density $\log_{10} f_{\rm rel}$ (3).}
\label{fig-mass-corr}
\end{figure*}

The distributions in Fig.~\ref{fig-hist} show a preference for the masses 
$m_\DM \sim 200$ GeV and  $m_A \sim m_\pm \sim 280$ GeV.  This is further 
illustrated by the scatter plots of Fig.~\ref{fig-mass-corr}, showing correlations of 
masses $m_\pm$ and $m_A$ and of $f_{\rm rel}$ versus $m_\DM$.  These reveal 
narrow islands of $m_\DM \sim m_h/2$ that correspond to the finely-tuned case of
resonant enhancement of DM annihilation with Higgs in the $s$ channel.
This allows the relic density to be suppressed despite having
$\lambda_\DM\ll 1$ in order to satisfy direct detection constraints,
and was the case focused upon in ref.\ \cite{Cline:2012hg}.  Here
we see that these regions now constitute only a small fraction of
our sample, which was enhanced by our choice of 
$(v_c/T_c)/\lambda_\DM$ rather than $(v_c/T_c)$ as the quantity to be
maximized by the MCMC.  There is a gap in the region $m_h < m_\DM <
m_W$ because the annihilation cross section below the threshold for W-pair 
production is too small to achieve a low enough relic density for these DM
masses. 
 
\begin{figure}[tb]
\centerline{$\!\!\!\!\!\!\!\!$
\includegraphics[width=0.48\textwidth]{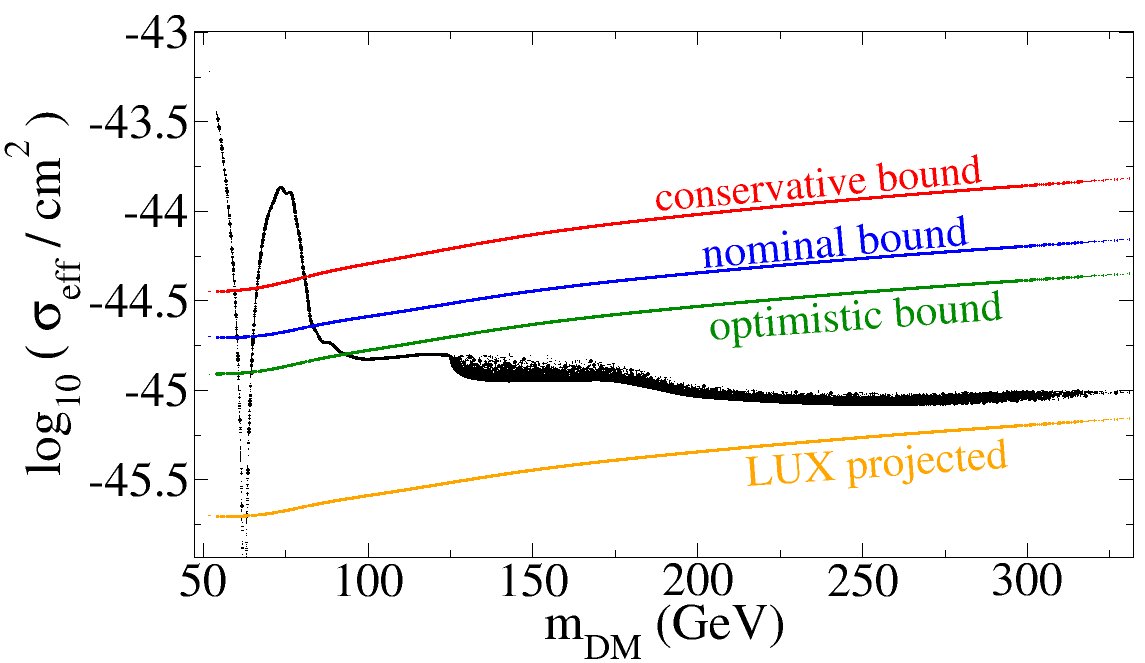}}
\vskip-0.3cm
\caption{The scatter of the expected cross section for direct detection
vs.\ $m_{\rm DM}$. Central curve (blue) shows the nominal 
XENON100 bound~\cite{Aprile:2012nq} and the upper and lower (red and green)
lines respectively are the more 
conservative/optimistic bounds reflecting uncertainties due to local DM 
distribution.  Bottom curve is expected constraint from LUX experiment.}
\label{fig-f-corr2}
\end{figure}

Fig.~\ref{fig-hist} shows that the
cutoff $\Lambda$ tends to be only a few TeV, indicating that this is
an effective theory which must be UV-completed by additional new
physics at a relatively low scale.  The enhancement in the phase
transition strength is modest, $v_c/T_c< 1.5$, but still sufficient
for electroweak baryogenesis. The ratio $f_{\rm rel}/f_{\rm Xe}$ tends to
be small, but can easily exceed 0.2, say, indicating that such models
will be discoverable by direct detection in XENON-like experiments
given a 5-fold increase in sensitivity.  This is also illustrated in 
Fig.~\ref{fig-f-corr2}, which shows the distribution of the accepted models in
the plane of $m_{\rm DM}$ versus the effective interaction cross section 
$\sigma_{\rm eff}\equiv f_{\rm rel}\sigma_{\rm SI}$, which takes into account
the reduced sensitivity of direct detection experiments due to the subdominant
nature of the dark matter.  The lower curve is our estimate of the bound which is
anticipated to be set by the LUX experiment \cite{Akerib:2012ys}; we
simply assumed that the $m_{\rm DM}$-dependence of the constraint is the
same as that of XENON100, scaled by anticipated LUX limit of 
$\sigma < 2\times 10^{-46}$cm$^2$ at $m_{\rm DM}=40$ GeV  
\cite{Fiorucci:2013yw}.  This is expected to be achieved in 2013; thus
our proposal will be tested over its entire relevant mass range in the very
near future.

As for the overall density of the inert DM component, the rightmost
graph of Fig.~\ref{fig-f-corr2} shows that only a few finely-tuned
models with $m_\DM$ close to $m_h$, or to the $m_W$ threshold, have 
$\Omega_\DM$ close to the observed value.  The majority of models
have $10^{-3} \lesssim f_{\rm rel} \lesssim 0.03$.

\begin{figure}[tb]
\vskip-0.2cm
\centerline{
\includegraphics[width=0.37\textwidth]{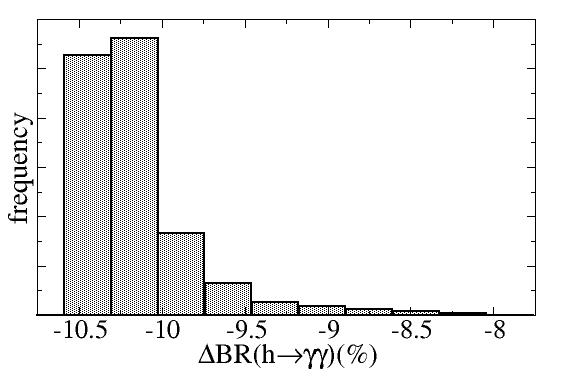}}
\vskip-0.3cm
\caption{Distribution of changes in the branching ratio for $h\to\gamma\gamma$
decays relative to the standard model prediction, for models generated
by the MCMC.}
\label{fig-deltaBR}
\end{figure}

In ref.\ \cite{Cline:2012hg} it was noted that due to contributions of
the charged Higgs, a 10\% decrease in the branching ratio for Higgs
decays to two photons, relative to the standard model expected value,
was predicted for the models of interest. There were experimental hints of
an excess in this channel from ATLAS \cite{ATLAS:2012gk} and CMS 
\cite{CMS:2012gu}, but these have become compatible with standard
model predictions in more recent data. We find that the earlier prediction of a 10\%
reduction is unchanged despite the larger region of parameter space available in
the present scenario. The distribution of the percent change in the
branching ratio for  $h\to\gamma\gamma$ is shown in Fig.~\ref{fig-deltaBR}. \\

\noindent{\bf Conclusions.} We have shown that it is in general easy to find
a strongly first order phase transition, desirable for electroweak
baryogenesis, in the context of inert doublet models that also provide some
fraction of the observed dark matter. The lightest inert
doublet state is found to typically  contribute $0.1-3\%$ of the total dark
matter. Its interactions with nucleons are nevertheless strong enough to be
directly detectable given a factor of 5 improvement over  the current
sensitivity of XENON100, which is expected to be achieved this year by LUX.
Moreover, we showed that the model predicts a 10\% decrease in the branching
ratio for Higgs decays to two photons.  We did not attempt to the compute
baryon asymmetry that might be achieved in this model.  To do so, one  should introduce a source of
CP-violation in the fermion sector, for example a dimension-6 operator
contributing to the
top quark mass, of the form $\eta|H|^2 \bar t H t$ with complex coupling $\eta$, 
and numerically evaluate the Boltzmann equation network for chemical 
potentials of chiral quark species near the electroweak bubble walls that form
during the phase transition.
Based on experience from similar studies
elsewhere~\cite{Cline:2011mm,Cline:2012hg}, it seems possible that 
the model augmented in this way could produce the observed baryon asymmetry 
of the universe. If dark matter is directly detected in the 100-300
GeV mass range, we will be encouraged to pursue this further.\\

\noindent {\bf Acknowledgements}.  We thank D.\ Borah for helpful correspondence
and earlier contributions to our code, and M.\ Trott for providing EWPD
constraints.  JC's research is supported by NSERC (Canada).

\bibliographystyle{apsrev}

\end{document}